\documentclass[11pt, a4paper]{article}    
\usepackage{amsmath,amssymb,latexsym}
\usepackage{graphicx}
\usepackage{bbm}
\usepackage{amsfonts}

\newcommand{\id}{\textrm{d}}
\newcommand{\mean}[1]{{\left< #1 \right>}}

\newcommand{\bbR}{{\mathbb R}}

\begin{document}

\title{Nonequilibrium linear response for Markov dynamics,\\
 I: jump processes and overdamped diffusions}

\author{ Marco Baiesi, Christian Maes, Bram Wynants$^*$\\
Instituut voor Theoretische Fysica, K.~U.~Leuven, B-3001 Leuven, Belgium\\
*{\tt bram.wynants@fys.kuleuven.be}
}
\date{\today}

\maketitle

\begin{abstract}
Systems out of equilibrium, in stationary as well as in
nonstationary regimes, display a linear response to energy
impulses simply expressed as the sum of two specific temporal
correlation functions. There is a natural interpretation of these
quantities. The first term corresponds to the correlation between
observable and excess entropy flux yielding a relation with energy
dissipation like in equilibrium.  The second term comes with a new
meaning: it is the correlation between the observable and the
excess in dynamical activity or
reactivity, playing an important role in dynamical fluctuation
theory out-of-equilibrium. It appears as a generalized escape rate
in the occupation statistics. The resulting response formula holds
for all observables
 and allows
direct numerical or experimental evaluation, for example in the
discussion of effective temperatures, as it only involves the
statistical averaging of explicit quantities, e.g. without needing
an expression for the nonequilibrium distribution. The physical
interpretation and the mathematical derivation are independent of
many details of the dynamics, but in this first part they are
restricted to Markov jump processes and overdamped diffusions.
\end{abstract}
keywords: nonequilibrium, fluctuation-dissipation, linear response.\\
PACS: 05.70.Ln, 05.20.-y


\section{Introduction and main result}\label{intro}
To know a system is to know its reaction to external stimuli.
Linear response theory applies when the effects are comparable to the causes and so far
has been systematized only for
systems in complete thermodynamic equilibrium \cite{mar,bala,mz}.
The result is summarized in the fluctuation-dissipation theorem
 \cite{ku,ca} stating that the response of an equilibrium
system to small external perturbations is proportional to its
fluctuations: an impulse changing the potential $U \rightarrow U -
h_s \,V$ at time $s$, will produce a response
$R_{QV}^{\textrm{eq}}(t,s)=\delta\mean{Q(t)}/\delta h_s|_{h=0}$ in
a quantity $Q$ at time $t\ge s$ given by
\begin{equation}\label{efd}
R_{QV}^{\textrm{eq}}(t,s) = \beta\,\frac{\id}{\id s}\mean{V(s)\,Q(t)}_{\textrm{eq}}
\end{equation}
Here $\mean{V(s)\,Q(t)}_{\textrm{eq}}$ is the equilibrium time
correlation function between $V(s)$ and $Q(t)$ and $\beta=1/k_BT$
is the inverse temperature.  For example, for an Ising spin system
$\sigma_j=\pm 1$ in equilibrium with energy function $U(\sigma)$,
where one looks
 for the response in the magnetization
$Q=\sum_j \sigma_j=V$ to a (time-dependent) magnetic field, one finds the (time-dependent)
susceptibility in terms of the spin-spin correlations in the equilibrium
process:
\[
R_{QV}^{\rm{eq}}(t,s) = {\cal R}(t-s) = \beta\,\sum_{i,j}
\frac{\id}{\id s}\langle
\sigma_i(s)\,\sigma_j(t)\rangle_{\rm{eq}}, \quad 0 < s < t
\]
When $h_s =h$  is constant over  $s\in [0,\tau]$, then to first
order in $h$ and for $t\geq \tau$, we get the thermoremanent
magnetization by integrating the previous formula:
\begin{equation}\label{chi}
\sum_{i} \big[\langle\sigma_i(t)\rangle^h -
\langle\sigma_i(0)\rangle_{\rm{eq}}\big] =
h\,\beta\,\sum_{i,j}\,\langle [\sigma_i(\tau)
-\sigma_i(0)]\,\sigma_j(t)\rangle_{\rm{eq}}
\end{equation}
The equilibrium process $\langle\cdot\rangle_{\rm{eq}}$ is
time-reversible and can for example be specified here by a Glauber
spin flip dynamics leaving the equilibrium distribution $\sim
\exp[ -\beta U(\sigma) ]$ invariant. The usual proof of
\eqref{efd}--\eqref{chi} proceeds by applying first-order
time-dependent perturbation theory and by inserting the
equilibrium  condition.  We review it in Appendix \ref{per} and we
indicate
 the problem with this approach for extending it to nonequilibrium conditions.

  In the present paper we generalize
   the equilibrium response formula \eqref{efd} to nonequilibrium regimes
(stationary as well as nonstationary ones).
 In this first part we treat Markov jump
processes and overdamped diffusions as simple nonequilibrium models ignoring inertial degrees
of freedom. In this context, formula \eqref{efd} changes to \eqref{form} below. A second part
   \cite{part2}
  following this paper will give extensions, in particular to
 underdamped diffusion processes, which are
 noisy Hamiltonian systems, and to general space-time local
 processes that are defined from their distribution on path-space.

To get a first understanding of the meaning of our generalization
it is instructive to view already \eqref{efd} not just as the
outcome of a perturbative calculation on the level of evolution
operators, but rather in its relation with
 dissipative effects, cf.~\cite{gg}. The point is that \eqref{efd} is a
correlation with the entropy flux associated to the perturbation.
To see it, we apply \eqref{efd} to write  the difference $\langle
Q(t)\rangle^h - \langle Q(t)\rangle_{\rm{eq}}$ between the
perturbed average and the original value as the time-integral
\begin{eqnarray}\label{calcu}
&&\beta \int_0^t \id s\,h_s\,\frac{\id}{\id s}\left< V(s)Q(t)
 \right>_{\rm{eq}} \nonumber\\&&= \beta\left<\left\{h_tV(t) - h_0V(0) -
 \int_0^t\id s\,\frac{\id }{\id s}{h}_s\,V(s)\right\}\,Q(t)\right>_{\rm{eq}}
 \end{eqnarray}
and we recognize the correlation of $Q(t)$
with the entropy flux
 \begin{equation}\label{ents}
\frac 1{k_BT} \left\{h_tV(t) - h_0V(0) -  \int_0^t\id s\,\frac{\id}{\id
s}h_s\;V(s)\right\}
\end{equation}
Indeed  $h_tV(t) - h_0V(0)$ is  the extra change of energy in the
environment due to the perturbation, and $\int_0^t\id
s\,\dot{h}_sV(s)$ is the work done on the system by the
perturbation. To be precise, the entropy flux here is an excess
flux of the perturbed process with respect to the unperturbed one
($\propto U(t) - U(0)$). This has an immediate analogue under
nonequilibrium conditions when we think of the entropy flux
\eqref{ents} as the excess with respect to whatever steady or
transient entropy production the system might already possess.

However, from dynamical fluctuation  theory we have learnt that
there is more than just dissipation or entropy production that
governs the fluctuations around nonequilibrium \cite{mn,mnw,mnw1}.
A novel quantity has appeared, which we call dynamical activity or
traffic, a measure for the system's nervosity or internal
reactivity. This dynamical activity is time-symmetric, in contrast
with the entropy production which leads time's arrow.  It was
first introduced in  \cite{t-symm,t-symm_2} and studied in the
context
 of nonequilibrium phase transitions
\cite{DA_4}, of  critical behavior \cite{DA_3}, for characterizing
the ratchet effect \cite{dm}, and for large deviation theory, e.g.
in \cite{DA_1,app,mn}. Close to equilibrium, entropy production
and dynamical activity merge, as can be understood from observing
that when approaching equilibrium the currents loose their
direction to become merely elements of (time-symmetric) traffic
between the states of the system. Further away from equilibrium,
dynamical activity and entropy production are really different and
play side by side in characterizing the nonequilibrium. Then,
formula \eqref{efd} splits in two separate terms, one entropic,
the other {\it frenetic}, to be explained below. That is made
visible in a generalized response formula, written here for the
nonequilibrium situations to be specified below: \eqref{efd} must
be changed into
\begin{equation}\label{form}
 R_{QV}^\mu(t,s) = \frac{\beta}{2}\frac{\id}{\id s}\left< V(s)Q(t)
 \right>_\mu - \frac{\beta}{2}\left<LV(s)\,Q(t)\right>_\mu
 \end{equation}
In \eqref{form} we have an arbitrary initial distribution $\mu$
for a given nonequilibrium process (over which we average on the
right-hand side). The first term is exactly like in equilibrium
(but with 1/2); the second term is the correlation with the
linearized dynamical activity (the correlation between $Q(t)$ and
the function $\beta\, LV$ at time $s$) which is new with respect
to the equilibrium \eqref{efd}. For its interpretation we restrict
us here first to jump processes. For the dynamical activity we
must then look at the escape rates, i.e.  at the frequencies by
which the  Markov jump process leaves a state $x$. Suppose the
process has rates $W(x,y)$ for the transitions $x\rightarrow y$.
The escape rate at state $x$ gets changed by the perturbation
$h_sV$. Its excess is
\begin{eqnarray}\label{exfre}
\sum_y
W(x,y)\left\{e^{\frac{\beta\,h_s}{2}[V(y)-V(x)]}-1\right\}
&\simeq&
 \frac{\beta\,h_s}{2}\, \sum_y W(x,y)[V(y)-V(x)] \nonumber\\
&\equiv&
\frac{\beta\,h_s}{2} (LV)(x)
\end{eqnarray}
to linear order in $h$, and $L$ is the backward generator of the jump process:
\[
LV(x) = \left.\frac{\id}{\id s}\right|_{s=0}\langle
V(s)\rangle_{x}
\]
In \eqref{form} we have abbreviated $LV(s) = LV(x_s)$ for $x_s$
the state at time $s$.
 More details will come
  in the next section, where the precise set-up and meaning will be given. Section
   \ref{exs} gives examples and
    makes visible the relative contribution of the
     two terms in \eqref{form}.  Most importantly the new formula is presented in
Section \ref{dyn} and \ref{fren} with its statistical mechanical
interpretation. Section \ref{efft} makes the relation with the
ambition of effective temperature. Section \ref{prev} still
connects our work with previous formulations, especially those in
\cite{ckp,lipp,lipp2,diez,ss1,gaw,ru}.  The Appendices give (A) explicit
calculations for illustrating the formul{\ae} of Section
\ref{efft}, (B) the response-formula for discrete time Markov
chains, useful for simulation purposes, and (C) the comparison
with formal perturbation theory and with the co-moving frame
interpretation of \cite{gaw}.

\section{What nonequilibrium?}

We consider open system first-order dynamics realized in Markov
jump and diffusion processes. These define probabilities on the
trajectories $\omega = (\omega_t, t\geq 0)$ of the system by
specifying the updating mechanism $\omega_t \rightarrow
\omega_{t+\id t}$. The state at each time is denoted by
$\omega_t=x$ and corresponds to a reduced description of the
universe, such as positions or elements of a discrete set of
configurations. Steady equilibrium reservoirs are integrated out
and replaced by effective external forces and noise. The
nonequilibrium condition can be imposed by external driving fields
or by installing mechanical displacements or chemical gradients at
the boundaries of the system but all external  driving is here
assumed to be time-independent.  Examples will come in Section
\ref{exs}; further details will not matter.

The updating is given in  terms of a {\em generator} $L$ in the
sense that for all single-time observables $f(x)$ (functions of
the configuration $x$ of the system) and for all initial
distributions $\mu$:
\begin{equation}\label{backward}
 \frac{\id}{\id t}\mean{f(\omega_t)}_{\mu} = \mean{(Lf)(\omega_t)}_{\mu}
\end{equation}
which is equivalent to saying that
\[
\mean{f(\omega_t)}_{\mu} = \int dx\, \mu(x)(e^{tL}f)(x) = \langle
e^{tL}f\rangle_\mu
\]
 In words, $e^{tL}$ ``pulls a function $f(x)$ back to the
time of the initial density $\mu(x)$.'' $L$ is therefore often
called the backward generator. More in general, for  any two
observables $f$ and $g$,  their correlations at times $0<t<s$
satisfy
\begin{equation}\label{1st}
\frac{\id}{\id s}\left< f(\omega_s)\,g(\omega_t) \right> = \left<
(Lf)(\omega_s)\,g(\omega_t) \right>,\quad 0 < t < s
\end{equation}
for an arbitrary average at time zero and over all allowed
trajectories.

We now specify the two classes of processes we deal with in this
first part.

\subsection{Jump processes}\label{jum}

We consider a finite
space $K$ of states $x,y,\ldots$ on which transition rates $W(x,y)$ give the probability per unit
time to jump between the states $x\rightarrow y$. No detailed balance is assumed.
We write $\rho$ for a stationary distribution, $\sum_{y\in K} [\rho(x)\,W(x,y) -
\rho(y)\,W(y,x)] = 0, \quad x\in K$. In this case the generator $L$ is
the matrix with off-diagonal elements $L_{xy} = W(x,y)$ and with
diagonal elements equal to minus the escape
rates, $L_{xx} = - \sum_y W(x,y)$. Hence
\[
Lf(x) = \sum_y W(x,y)\,[f(y) - f(x)]
\]
and by stationarity $\sum_x \rho(x) \,Lf(x) =0$.

We think of  the states $x$ as configurations of a mesoscopic
system undergoing a Markov evolution, such as for chemical
kinetics or for models of interacting
components on
a lattice which are driven away from equilibrium, cf. example \ref{ex1}; the state $x$ is then the
total configuration and the transitions are local; response in discrete time is treated in
Appendix \ref{disc}.

At time $t=0$ we draw the initial state from an arbitrary probability
distribution $\mu(x), x\in K$. For $t>0$ we apply the perturbed dynamics with transition
rates
\begin{equation}\label{jw}
W_t(x,y) = W(x,y)\,e^{\frac{\beta\, h_t}{2} [V(y) - V(x)]}
\end{equation}
for some potential $V$ with small amplitudes $h$. We thus switch on another channel of energy
exchange with a reservoir at temperature $T$. This is a standard type of perturbation
in Markov jump processes by which we add a potential; we will stick to that. More general types of perturbation are possible, as
for
example in \cite{diez} but for these the specific formula (\ref{form}) has to be modified, see \cite{mprf}.

\subsection{Overdamped diffusions}\label{overd}

Overdamped diffusions are stochastic processes
defined in the It\^o-sense by
\begin{equation}\label{gsd}
  \id x_t = \bigl\{\nu(x_t) \bigl[ F(x_t) -
   \nabla U(x_t) \bigr] + \nabla\cdot D(x_t)\bigr\}\,\id t
  + \sqrt{2 D(x_t)}\, \id B_t
\end{equation}
Now $x\in \bbR^d$ and the noise is present in the form of
the $d-$dimensional vector $\id B_t$ having independent
standard Gaussian white noise components. We assume that the environment is in thermal
equilibrium at inverse
temperature $\beta>0$ imposing the condition $\nu(x) =
\beta D(x)$ between the bare mobility $\nu$ and the diffusion matrix $D$; they are strictly
 positive (symmetric) $d
\times d-$matrices.  The relation $\nu=\beta\, D$ is easily
confused with the fluctuation-dissipation theorem; they are only
equivalent in equilibrium.  More in general, $\nu=\beta\, D$ is an
expression of local detailed balance assuring the proper physical
identification of the various terms in the equation \eqref{gsd}.
 There is a potential $U$ and the force $F$ represents the
nonequilibrium driving.  We do not specify here regularity
properties or boundary conditions. Equilibrium \eqref{efd} is
obtained when $F=0$, for stationary $\rho \propto \exp( -\beta U
)$. The diffusions \eqref{gsd} are called {\it overdamped} because
they have forces proportional to velocities. These processes are
high damping limits (sometimes called
 Smoluchowski limits) of underdamped or inertial stochastic dynamics that will be
considered in Part II \cite{part2}. On the other hand, formal
extensions to stochastic and driven Ginzburg-Landau models (models
A-B-C in \cite{gamb}) do not seem difficult. Examples come in
Sections \ref{ex2} -- \ref{ex3}.

  The generator of the dynamics \eqref{gsd} reads
\begin{equation}\label{geov}
L =
\nu(F-\nabla U)\cdot \nabla + \nabla D\cdot\nabla
\end{equation}
 The time-dependent perturbation changes $U$ in \eqref{gsd} into
 $U-h_t\,V$ for $t\geq 0$.  In other words the Hamiltonian part of the
dynamics of the system gets an additional  conservative force with
a time-dependent amplitude.  Here also we assume that $\nu$ and $D$ are unchanged
under adding the potential $h_tV$.

\subsection{Main question}
What to expect at time $t>0$ for an observable
$Q$ ? There will be a shift in its value, deviating both from the original expectation and from the
stationary value:
\begin{eqnarray}
\langle Q(t)\rangle^h_\mu &\neq& \langle Q(t)\rangle_\mu\nonumber\\
&\neq& \langle Q(t)\rangle_\rho = \langle Q(0)\rangle_\rho
\end{eqnarray}
Note that we  abbreviate $Q(t) =  Q(x_t)$.  The right-hand
sides average over the unperturbed dynamics, the upper one starting from $\mu$
and, on the next line, when starting from the stationary $\rho$.
Linear response theory for systems out of equilibrium aims at estimating and interpreting the
deviations
\[
\langle Q(t)\rangle^h_\mu - \langle Q(t)\rangle_\mu
\]
to first order in $h$.   In this linear regime, one studies the
deviations of an observable $\mean{Q(t)}^h$ from its expected
value $\mean{Q(t)}$ in the unperturbed dynamics,
\begin{equation}\label{linres}
\mean{Q(t)}^h = \mean{Q(t)} + \int_0^tds\, h_s R_{QV}(t,s)
\end{equation}
where
\begin{equation}\label{R_QV}
R_{QV}(t,s)\equiv \frac{\delta\mean{Q(t)}^h}{\delta h_s}\Big|_{h=0}
\end{equation}
is the response of a quantity $Q(t)$ to a (small) impulse $h_s$ at
a previous time $s<t$.

The initial distribution $\mu$ will be remembered via super- and
subscripts, where appropriate.  A special and interesting
case is the stationary response for $\mu=\rho$.

\section{Nonequilibrium response formula}

Under the previous set-up there is a simple general formula for
the linear response derived in \cite{prl}, the same for all observables $Q$ and $V$:
\begin{equation}\label{form1}
 R_{QV}^\mu(t,s) = \frac{\beta}{2}\frac{d}{d s}\left< V(s)Q(t)
 \right>_\mu - \frac{\beta}{2}\left<LV(s)\,Q(t)\right>_\mu
 \end{equation}
The first example of a formula of this kind is Eq.(22) in Ref.~\cite{lipp}, dealing
with discrete spin systems.
Since random paths are not smooth, it does not make sense to take the time-derivative
inside the expectation in the first term.
On the other hand, for $s > t$, \eqref{1st} can be
applied and as a consequence causality is automatically verified
in \eqref{form1}: the response vanishes for $s>t$.  One can thus
also rewrite \eqref{form1}  in a more symmetric
way which is however only valid for $t\geq s$:
\begin{equation}\label{asymm}
 R_{QV}^\mu(t,s) = \frac{\beta}{2}
\left[
\frac{\id }{\id s}\big<V(s)Q(t)\big>_\mu-\frac{\id}{\id
t}\big<V(t)Q(s)\big>_\mu
\right]
-\frac{\beta}{2} \big<LV(s)Q(t)-LV(t)Q(s)\big>_\mu
\end{equation}
Equations of this form can also be found back in the literature~cite{ckp,lipp,diez}.
  {\it Our
main goal here is to extend these earlier results and others (such
as in \cite{gaw,jona}) into the general and usable formula
\eqref{form1}, and, most importantly, to accompany it with an
interpretation derived from dynamical fluctuation theory.}  That
will be continued in \cite{part2}, showing further robustness of
the interpretation.

In the introduction Section \ref{intro} we have already  briefly
introduced the statistical meaning of the two terms
on the right-hand side of \eqref{form}=\eqref{form1}. More is to come in Sections \ref{dyn} and
\ref{fren}.
The first term in \eqref{form1} (as in \eqref{efd}) is a
correlation with the excess entropy flux.
Secondly, there is the correlation between $Q(t)$ and $\beta LV(s)$,
a quantity that we call {\em frenesy},
derived from the adjective frenetic or frantic.   In contrast
to the entropy production, which has a preferred direction in time,
frenesy is a time-symmetric quantity.

Let us also see how formula \eqref{form1} reconstructs
  the equilibrium formula \eqref{efd}.
In this case $\mu=\rho$ is the equilibrium distribution with the process
satisfying time-reversal symmetry, with $s < t$
\begin{eqnarray}\label{eqfren} && \langle
LV(s) \,Q(t)\rangle_{\rm{eq}} =
 \langle LV(t) \,Q(s)\rangle_{\rm{eq}}\nonumber\\&&
=  \frac{\id}{\id t}\langle V(t) \,Q(s)\rangle_{\rm{eq}} =
- \frac{\id}{\id s}\langle V(t) \,Q(s)\rangle_{\rm{eq}}= - \frac{\id}{\id s}\langle
V(s)
\,Q(t)\rangle_{\rm{eq}}
\end{eqnarray}
which is the first term in \eqref{form1}. Since by \eqref{eqfren} in
equilibrium
the frenesy-correlation exactly
equals minus the correlation with the entropy flux,
the two terms on the right-hand side of \eqref{form1}
add up to give \eqref{efd}.

We end the section with the formal proof of formula \eqref{form1},
which is easy when going to path-space. We should write the
perturbed expectation value in terms of the unperturbed one. More
explicitly we consider paths $\omega = (\omega_s)$ over the
time-interval $s\in[0,t]$ to write
\[
\langle Q(t)\rangle^h = \int\,\id P_{\mu}(\omega)\,  \frac{\id
P^h_\mu}{\id P_\mu}(\omega) \,Q(\omega_t)
\]
where the $\id P$'s stand  for path-probability densities in the
perturbed (superscript $h$) and unperturbed processes.  In
particular, for the jump  processes of Section \ref{jum},
\[
P_\mu(\omega) =
\mu(\omega_0)W(\omega_0,\omega_{t_1})e^{-\sum_y\int_0^{t_1}\id s W(\omega_s,y)}\ldots
W(\omega_{t_{n-1}},\omega_{t_n})e^{-\sum_y\int_{t_n}^{t}\id s W(\omega_{s},y)}
\]
for a path $\omega=(\omega_0\rightarrow \omega_{t_1}\rightarrow\ldots\rightarrow
\omega_{t_n} =\omega_t)$.
As a consequence,
\begin{eqnarray}\label{pro}
\log \frac{\id P^h_\mu}{\id P_\mu}(\omega)
&&= \frac{\beta}{2}\sum_{k=1}^n\,h_{t_k}\,\big[V(\omega_{t_k})-
V(\omega_{t_{k-1}})\big]\nonumber\\
- && \sum_y\int_0^t\id s \,W(\omega_s,y)\big[e^{\frac{\beta
h_s}{2}[V(y) - V(\omega_s)]}-1\big]
\end{eqnarray}
where the first sum is over all the jump times
$(t_1,\ldots,t_n)$ in $\omega$, and we put $t_0=0$ and $t_{n+1}=t$. Remember that
 the path is constant between the jump times so that
\begin{eqnarray}
\int_0^t \id s\,\frac{\id}{\id s} h_s \; V(\omega_s) &=&
\sum_{k=0}^n V(\omega_{t_{k}}) \, \big[ h_{t_{k+1}} - h_{t_{k}}\big ]
\nonumber\\
= h_t V(\omega_t) - h_0 V(\omega_0) & + &
\sum_{k=1}^n\,h_{t_k}\,\big[V(\omega_{t_{k-1}})- V(\omega_{t_{k}})\big] \nonumber
\end{eqnarray}
by partial summation.  We can therefore rewrite the first line of
\eqref{pro} and substitute \eqref{ents}. The rest is expansion to
first order in $h$ for a finite number of terms, almost surely
under $\id P_{\mu}(\omega)$.  In particular, the frenetic term appears as the first
order in the time-symmetric part (second line of \eqref{pro}), like in \eqref{exfre}.

The strategy is
unchanged for overdamped diffusions. There, the paths $\omega = (x_s)$ are continuous and the
action
\eqref{pro} can be found in the stochastic integral
\begin{eqnarray}\label{proov}
\log \frac{\id P^h_\mu}{\id P_\mu}(\omega)
&&= \frac{\beta}{2}\Bigg\{\int_0^t \id x_s \,\nabla V(x_s)\, h_s\nonumber\\
+&& \int_0^t \id s\,\nu(\nabla U - F)(x_s)\cdot h_s \nabla V(x_s) + O(h^2)\Bigg\}
\end{eqnarray}
The first It\^o-integral can be rewritten in the Stratonovich
sense (with the ``$\circ$''-notation in the integral)
\begin{eqnarray}
\int_0^t \id x_s \nabla V(x_s)\, h_s & = &\int_0^t \id x_s \circ \nabla V(x_s)\, h_s
\nonumber\\
&&+ \int_0^t\id s h_s\nabla D\cdot\nabla V(x_s)
\end{eqnarray}
 where the last term can be combined with the second line in \eqref{proov} to make
  $\beta LV /2$, see \eqref{geov}. Moreover,
 \[
\left< \int_0^t \id x_s \circ \nabla V(x_s)\, h_s \,Q(x_t)\right>_\mu =
\int_0^t \id s \, h_s\,\frac{\id}{\id s}\langle V(s)Q(t)\rangle_\mu
\]
  The rest is again trivial expansion.

\section{Examples}\label{exs}
It is often more convenient to visualize the integrated version of \eqref{form1}
for a small but constant perturbation $h_s = h, s \ge 0$.  Then, the generalized
susceptibility
\[
\chi(t) = \frac 1{h}\left[ \,\mean{Q(t)}^h - \mean{Q(t)} \right] \;,
\]
is given by
\begin{equation}\label{chit}
\chi(t) / \beta = \frac{1}{2}[ C(t) + K(t) ] \equiv C_{NE}(t)
\end{equation}
with correlation function (coming from the entropic term in
\eqref{form1})
\[
C(t) = \mean{V(t)Q(t)} -  \mean{V(0)Q(t)}
\]
and a term (coming from the frenetic term in \eqref{form1}, extra with respect to equilibrium)
\[
K(t) = - \int_0^t ds \mean{ LV(s)Q(t) }
\]
representing an integrated correlation function. The average of $C$ and $K$, denoted by $C_{NE}(t)$
in \eqref{chit}, thus has to be equal to $\chi(t) / \beta$ in
general. If extended to $t\uparrow \infty$, $\chi(t)$ gives the change
in nonequilibrium stationary expectation when adding a small potential.

\subsection{Driven Kawasaki dynamics}\label{ex1}

Consider an exclusion process as a model of ionic transport through a narrow channel.
 This is described
by a collection of $n$ sites, labelled by $i=1,\ldots,n$, each holding
either one particle ($x^i=1$) or none ($x^i=0$).
In the bulk of this system no particles are
created or annihilated, only jumping to neighboring sites is
allowed via a Kawasaki dynamics.
At the edges $i=1,n$ particles can move in or out from reservoirs
with density $d_1$ and $d_n$, respectively.
The form of the response formula \eqref{form1} is unchanged by adding a nearest neighbor
interaction with energy $U(x) = -\sum_{i=1}^nx^i x^{i+1}$.
Moreover, we can add an ``electric'' field $E$ promoting particle jumps to the right.
Since it is a Markov jump process, there are transition rates
for particles hopping to neighboring sites and rates for creation and
annihilation at the edges.
For example, a particle enters into site $i=1$ from the reservoir with rate
\[
W(x,y) = d_1 \psi(x,y)\,\exp\left\{ -\frac{\beta}{2} [U(y) - U(x)]\right\},\quad
\psi(x,y)=\psi(y,x)
\]
where $y=x$ except that $x^1=0$ while $y^1=1$.
The rate of the reverse transition can be found by imposing local detailed balance.
Similarly, in the bulk, when e.g.  $x^i=1, x^{i+1}=0$ and $y$ is reached by inverting these occupations,
\[
W(x,y) = \psi(x,y)\,\exp\left\{ -\frac{\beta}{2} [U(y) - U(x) - E]\right\},\quad
\psi(x,y)=\psi(y,x)
\]
Simple nonequilibrium conditions can thus be introduced either by
\begin{itemize}
\item[i)] setting different reservoir densities $d_1 \ne d_n$, or
\item[ii)] setting a nonzero electric field $E>0$ in the bulk.
\end{itemize}

 We choose the
total number of particles ${\cal N}(t)=\sum_{i=1}^n x^i$ as observable. We also introduce
a perturbation $V(s)$ equal to ${\cal N}(s)$, which means
that we are changing  the chemical potential of both reservoirs
with a common shift.
Transition rates for the perturbed process are thus multiplied by
a factor $e^{\beta h_s/2}$ if a particle enters the system, and by $e^{-\beta h_s/2}$ when a
particle leaves;
transitions in the bulk are left unchanged.

In this case the frenetic term is with $LV= {\cal J}(s)$, the  systematic
current, and represents the expected change of the potential ${\cal N}$ per unit time,
i.e., it is the rate of change in the number of particles from the two possible
transitions at boundary sites.
Thus, for $x\to y$ the transition modifying $x^1$, and for $x\to z$ the transition
modifying $x^n$, we have
\[
{\cal J}(x)\! =\!
[{\cal N}(y) - {\cal N}(x)] W(x,y)+
[{\cal N}(z) - {\cal N}(x)] W(x,z)
\]
We have numerically verified
that $\chi = C = K$ under equilibrium conditions. While $C = K$ to excellent
precision, the shape of $\chi$ depends weakly on $h$, and is found to
converge to $C$ only for $h$ sufficiently small. In fact, one can pretend
exact matching only in the limit $h\to 0$, but $h=0.01$ turns out to be
sufficiently small to achieve a good convergence.

\begin{figure}[!bt]
\begin{center}
\includegraphics[angle=0,width=10cm]{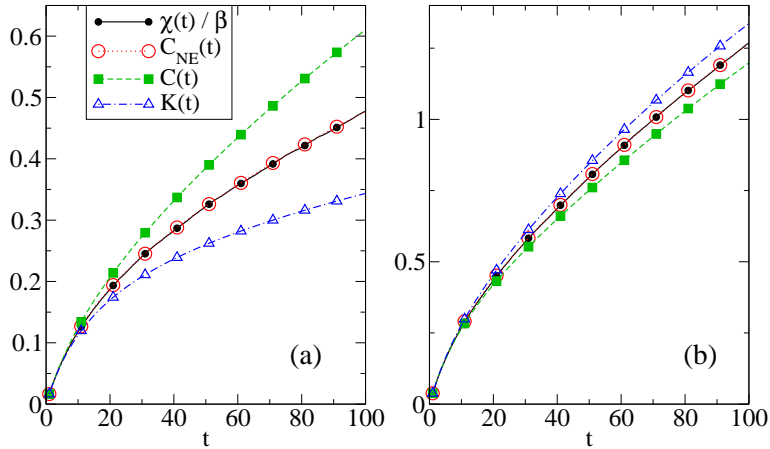}
\end{center}
\caption{Plot of the quantities involved in
Eq.~(\ref{chit}), for (a) case i) ($E=0$) with $L=10$, $\beta=1$,
$h=-0.01$, and reservoir density unbalance $d_1=0.9$, $d_n=0.1$,
and (b)  for case ii) ($d_1=d_n=0.5$) with $L=10$,
$\beta=1$, $h=-0.01$, $E=3$.
\label{fig:1}}
\end{figure}

A representative example for the nonequilibrium case i) is shown in
Fig.~\ref{fig:1}(a). One can see that each of the functions $C(t)$  and $K(t)$
is a poor approximation of the response, while
the agreement between $C_{NE}(t)$ and $\chi(t) / \beta$ is excellent.

In Fig.~\ref{fig:1}(b) we show an example of nonequilibrium condition ii).
Again, only $C_{NE}(t)$ matches $\chi(t) / \beta$.
Curiously, a comparison of this example with the previous one reveals that
$C$ can be either larger or smaller than $K$, even for two nonequilibrium conditions
that look pretty similar, in the sense that they yield a current in the
same direction for a relatively simple system.

\subsection{Diffusion on the circle}\label{ex2}

Consider the Langevin equation for a particle position $x_t \in S^1$ on a circle driven by a
constant field $F(x) = f$ and subject to a time-dependent forcing:
\begin{equation}
 \id x_t = \nu[f-U'(x_t)-h_tV'(x_t)]\id t + \sqrt{2D}\,\id B_t
\end{equation}
Here the prime denotes differentiation with respect to space. The
term $h_t\,V'$ is the time-dependent perturbation while $\id B_t$
is a standard Gaussian white noise.  The diffusion constant $D$
and the mobility $\nu$ are related by the Einstein relation $\nu =
\beta D$,  with $\beta$ the inverse temperature.

This example has been recently experimentally realized as reported in \cite{cili},
for testing the co-moving frame interpretation of \cite{gaw} also explained in Appendix
\ref{per}.
In fact, its nonequilibrium stationary distribution $\rho$ is known analytically, see
\cite{mnw1}.

\begin{figure}[!bt]
\begin{center}
\includegraphics[angle=0,width=10cm]{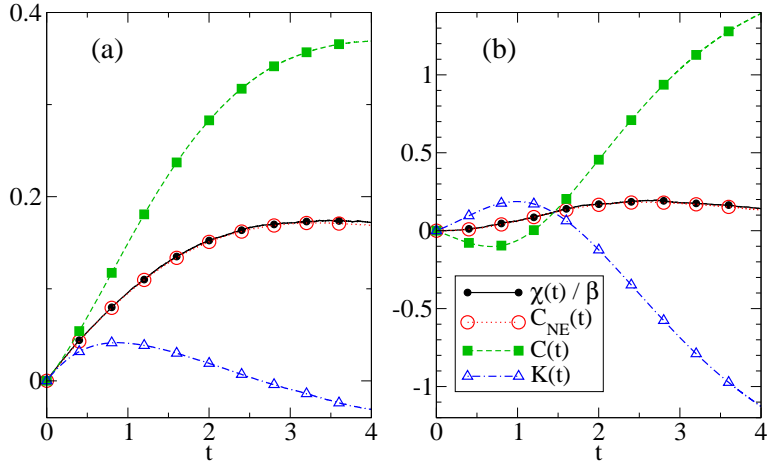}
\end{center}
\caption{Response and fluctuations of the overdamped particle
in a tilted periodic potential, as discussed in the text,
with inverse temperature $\beta=0.2$,
mobility $\nu=1$  and perturbation $h=-0.02$.
(a) Steady state regime, with initial distribution equal to the
stationary one, and force $f=0.9$.
(b) Transient regime, with initial position $x_0 = 0$ and force $f=0$.
}
\label{fig:overd}
\end{figure}

 Here we  show the result of a simple simulation of the overdamped
particle studied in the experiment, with  potentials and
observable $U(x)=Q(x)=V(x) = \sin(x)$, and a constant force $f$.
The frenesy $\beta L V(x)$ of a particle at position $x$
 can be computed by applying the generator
of the overdamped dynamics
\[
L = (f-U')\frac{\id}{\id x} + \frac{1}{\beta}\frac{\id^2}{\id x^2}
\]
to the potential $V(x)$, which gives $\beta L V(x) =
\beta[f-\cos(x)]\cos(x) - \sin(x)$. We have measured the
correlations $C(t)$ and $K(t)$ as well as responses
$\chi(t)$ for small $h$, see \eqref{chit}.

Fig.~\ref{fig:overd}(a) shows that for strong stationary nonequilibrium
$f\gg 0$ there is a large difference between the integrated
correlation function of the frenesy and that of the entropy
production. However, their average $C_{NE}(t)$ agrees very well
with the response $\chi(t)$, as it should.

It is important however to recall that our approach works for
nonstationary regimes as well. In Fig.~\ref{fig:overd}(b) we show
an example of a particle  starting at time $t=0$ from
a nonstationary initial distribution $\mu(x) = \delta_{x,0}$
(i.e. its position is $x_0=0$), but for the rest
not forced outside equilibrium, $f=0$,
to emphasize the transient character of this situation.
Again, we can see that the response is well estimated by $C_{NE}(t)$.

In the second part \cite{part2} we will treat the inertial version
of this model.

\subsection{Driven Brownian motion}\label{ex3}
Consider again the overdamped Langevin equation \eqref{gsd} but with constant $\nu = \beta D$:
\[
 (\id x^i_t) = \nu^{ij}[f^j(x_t) + h^j]\id t +\sqrt{2D}^{ij}\id B^j_t
\]
with repeated indices $j=1,\ldots d$ summed over.
The $f$ includes all forces and $h$ is the small constant perturbation.
    The true mobility $M$ is defined as the response
\begin{equation}\label{mmm}
 M^{ij} = \lim_{t\to\infty}\left.\frac{d}{dh^j}\left<
 \dot{x}^i_t \right>^h_{\rho}\right|_{h=0}
\end{equation}
while the real diffusion matrix is given by
\[
 \mathcal{D}^{ij} = \lim_{t\to\infty}\frac{1}{2t}\int_0^t\id s\int_0^t\id r\
\left<(\dot{x}^i_{s}-\left<\dot{x}^i\right>_{\rho})(\dot{x}^j_{r}-\left<\dot{x}^j\right>_{\rho}) \right>_{\rho}
\]
As \eqref{mmm} is a response function, we can easily compute it
within our framework. After a straightforward calculation, it
follows that
\begin{equation}\label{er}
 M^{ij} = \beta\mathcal{D}^{ij}
 -\frac{\beta\nu^{jk}}{2}\lim_{t\to\infty}\frac{1}{t}\int_0^t\id
 s\int_0^t\id
 r\left< f^k(x_{s})(\dot{x}^i_{r}-\left<\dot{x}^i\right>_{\rho}) \right>_{\rho}
\end{equation}
Note also that $\left<\dot{x}^i\right>_{\rho} = \left<\nu^{ij}f^j(x)\right>_{\rho}$.
Observe that in equilibrium, the second term in \eqref{er}
vanishes because the observable $\dot{x}_t$ is anti-symmetric, and $\left<\dot{x}\right>_{\rm{eq}}=0$. Then,
the equilibrium fluctuation-dissipation relation holds with
$M=\beta {\cal D}$.  Its violation for nonequilibrium, i.e., that
there is a correction (= second term) in \eqref{er}, is completely
compatible with the condition $\nu=\beta D$ for the reservoir in
thermal equilibrium at inverse temperature $\beta$.

\section{Dissipation: the entropic term}\label{dyn}
Even at equilibrium,
the response to an external perturbation is a nonequilibrium
process involving a transient with production of entropy.
In steady nonequilibrium conditions instead
we deal with the {\em excess} in this entropy production,
as the starting point is already a regime with nonvanishing flows.

In equilibrium it is known that the response function is
closely related to the energy dissipation of the system into the
environment. In systems driven out of equilibrium things get more
complicated: even when the system is not perturbed there is
already a (time extensive) heat dissipation. This is the
housekeeping heat that is needed to maintain the (unperturbed)
nonequilibrium stationary state. Perturbing the system then gives additional
heat. We see here that the prediction in~\cite{house} that the
usual equilibrium relation between response and dissipation is
preserved when taking into account only the {\it excess} heating
and ignoring the
housekeeping heat, is indeed true in the following sense.

During a specific trajectory of the system, the heat dissipation
can be split into two parts: ${\cal Q}_{hk}+{\cal
Q}_{ex}$~\cite{house,house2}. The first term is the housekeeping
heat while ${\cal Q}_{ex}$ is the extra heat generated through the
perturbation, and is just the temperature times the excess entropy
production, which we already encountered in our calculations
\eqref{calcu}--\eqref{ents}.

 Let us assume that the amplitude
$h_s$ is periodic with period $t$ so that it makes sense to
consider the expectation of the  excess heat over the interval
$[0,t]$:
\[
 \left< {\cal Q}_{ex} \right>_{\rho}^h
=\int_0^t\id s\,h_s\,\frac{\id}{\id s}\left< V(s) \right>_{\rho}^h
\]
We expand the expectation value of $V$ using the definition of the
response function $R_{VV}$:
\[
\left< {\cal Q}_{ex} \right>_{\rho}^h =\int_0^t\id
s\,h_s\,\frac{\id}{\id s}\left[\left< V(s) \right>_{\rho} +
\int_0^s\id r\,R_{VV}(s,r)h_r\right]
\]
Here $\left< V(s) \right>_{\rho}$ is independent of $s$ because of
stationarity and $R_{VV}(s,r)={\cal R}(s-r)=0$ for $s<r$ so that
\[
 \left< {\cal Q}_{ex} \right>_{\rho}^h
=-\int_{0}^t\id s\,\frac{\id}{\id
s}h_s\,\int_{-\infty}^{+\infty}\id r\,{\cal R}(r)h_{s-r}
\]
where the right-hand side is exactly the same expression as in
equilibrium for the dissipated energy over one period.
As is standard, see e.g. \cite{mz,mar}, the last relation can be
rewritten in the frequency domain to obtain now the (excess) heat
proportional to the imaginary part of the Fourier transform of
$R_{VV}={\cal R}$. We conclude that also out-of-equilibrium the
imaginary part of the Fourier transform of the generalized
susceptibility is related to energy dissipation, but only of the
excess heat. Observe finally that the imaginary part of the
Fourier transform is the Fourier transform of the
time-antisymmetric part of $R_{VV}$ and that we have already
written down this time-antisymmetric part: it is the right-hand
side of \eqref{asymm}, extended  to $t\le s$.

\section{Activity: the frenetic term}\label{fren}

As we have seen,
besides the entropy flux there is yet another
relevant quantity getting into the picture.  It is a younger
concept in nonequilibrium studies as witnessed by the fact that
its name has not been settled yet: it has been called {\em
traffic}~\cite{mn,mnw1,mnw}  or dynamical {\em
  activity}~\cite{DA_1,app,DA_3,DA_4}, to denote a property related
with the volume of transitions or changes performed in time. We
like to refer to it as frenesy, a name that sounds
more similar to entropy or energy. This quantity is
time-symmetric, in the sense that trajectories display the same
activity if they are spanned in the normal temporal direction or
backward in time.  The relevance of time-symmetric quantities has
been anticipated in \cite{t-symm,t-symm_2}.  Activity denotes thus
an aspect of the dynamics that complements entropy production
(which is time anti-symmetric -- flows reverse together with time)
in the description of fluctuating quantities in regimes out of
equilibrium.

We recall here how the frenesy/dynamical activity
appears in dynamical fluctuation theory, choosing the set-up of Section
\ref{jum} (finite state space Markov
jump processes $x_t$).

Consider the stationary process $P_\rho$ and measure the fraction
of time that the system spends in each state $x\in K$:
\[
p_\tau(x) = \frac 1{\tau}\,\int_0^\tau \delta_{x_t,x}\,\id
t,\quad \mbox{with}\; \delta_{a,b} = 0 \mbox{ if } a\neq b \;\mbox{and}\; \delta_{a,b}= 1 \mbox{
if } a = b
\]
That is the empirical distribution of occupation times over
$[0,\tau]$ while drawn at time zero from the stationary $\rho$.
When the stationary
process $P_\rho$ is an ergodic Markov process, $p_\tau$ is invariant under time-reversal and
$p_\tau
\rightarrow \rho$ for $ \tau \uparrow +\infty$. Following the
pioneering work in \cite{DV} we look here at the fluctuations
around that law of large times, for which it is known that
\[
\mbox{Prob}_\rho[p_\tau\simeq \mu] \simeq e^{-\tau I(\mu)},\quad \tau \uparrow +\infty
\]
See e.g. \cite{DZ} for a precise description of this fluctuation formula.  The
exponent is governed by the functional $I(\mu)\geq 0$, with equality
for $I(\rho)=0$.  There is a variational expression for
the fluctuation functional $I(\mu)$,
\begin{equation}\label{dv}
I(\mu) = \sup_{V}
\left\{
-\sum_x \mu(x) \left[\sum_y W(x,y) e^{\frac{\beta}{2} [V(y)-V(x)]} - \sum_y W(x,y)\right]
\right\}
 \end{equation}
 in terms of the excess escape rates when adding a potential $-V$ much as in
\eqref{jw}. The actual supremum is then reached for the potential $V$ that makes $\mu$
stationary; i.e., by changing
$W(x,y)\rightarrow W(x,y)\,\exp\{ \beta[V(y) - V(x)]/2 \}$. Thus, for that $\mu$-dependent
potential $V$,
   \begin{eqnarray*}
 I(\mu) &=& \sum_{x,y}
  \mu(x)\, W(x,y) - \sum_y \mu(x)\, W(x,y)\,e^{\beta[V(y) - V(x)]/2}\nonumber
 \\ &\simeq &
-\frac{\beta}{2}\sum_x\mu(x)\,LV(x)
\end{eqnarray*}
In the second  line (expected frenesy in $\mu$) we have assumed that $\mu$ is
close to the stationary measure (small fluctuations), and we have
written the first term in an expansion around $\mu -\rho$, where
$V$ is supposed to be small as well, again like in \eqref{exfre}. In other words, the rate
$I(\mu)$ of escape from density $\mu$ is given in terms of the expected frenesy under $\mu$.

For further information on the role of activity/frenesy in dynamical
fluctuation theory, we refer to \cite{mn,mnw,mnw1}.

\section{To effective temperature}\label{efft}

While the equilibrium fluctuation--dissipation relation
\eqref{efd} is typically violated for nonequilibrium regimes, one
may wonder whether it can sometimes be restored by the
introduction of an effective temperature $T^{\rm{eff}}$, in the
sense
\begin{equation}\label{ef}
R_{QV}^\mu(t,s)= \frac 1{k_BT^{\rm{eff}}}\,\frac{\id }{\id s}
\langle V(s) Q(t)\rangle_\mu
\end{equation}
Many studies have been devoted to the study of the prefactor, in what sense it perhaps
resembles a
   thermodynamic temperature-like quantity for some classes of
   observables and over some
    scales of times $(s/t,s)$; we refer to \cite{kur,cr,gamb,henk} for an
    entry into the extensive literature.  Clearly, whatever the
         purpose of the discussion, an exact expression of the response should
          help, especially when entirely
          in terms of explicit correlation functions.  The first calculations
in this sense are in \cite{Xa} and they have been referred to as the ``no field-method''
\cite{ric}. In particular, for purposes of simulation or numerical verification of \eqref{ef}
we
do no longer need to perform the perturbation by hand.
In fact, now we can write the ratio $T/T^{\rm{eff}}= X$ entirely
in terms of correlation
 functions
 \begin{equation}\label{x} X= X_{QV}(\mu;t,s) = \frac 1{2}\big[ 1 - \frac{\langle
LV(s)\,Q(t)\rangle_\mu}{\frac{\id}{\id s} \langle V(s)
Q(t)\rangle_\mu}\big]
\end{equation}
with numerator and denominator in \eqref{x} each having a specific
physical meaning as in the previous two sections. An effective
temperature is obtained as the ratio between the frenetic and the
entropic term: if for some observables $(V,Q)$ and over
time-scales
 $(t/s,t)$,
\[
Y\;\frac{\id}{\id s}\langle
V(s)\,Q(t)\rangle_\mu = \langle LV(s)
\,Q(t)\rangle_\mu
\]
for some $Y$, then $X=(1-Y)/2$.  Equilibrium has   $X=1=-Y$.  In
the case where $LV\approx 0$ as for a conserved quantity, then
$Y=0$. The simplest example is Brownian motion $L=\Delta$ for $V$ the position, cf.
Virasoro's example in
\cite{ckp}. In that last reference, what
are called ``flat directions'' can be associated to perturbations with zero frenesy
$LV=0$.
In the Appendix \ref{gau} we give some explicit results to illustrate the above ideas
   and formul{\ae}.

  Finally,  $X$ and the effective temperature
$T^{\rm{eff}}$ get negative when the frenetic term overwhelms the
entropic contribution, which we can expect whenever the
perturbation strongly activates the dynamics. It is not clear to
us whether that relates with the observations in \cite{ru}
concerning ``activated dynamics,'' but certainly, those ``active
states'' with negative effective temperature, here interpreted as
highly frenetic ones are very different from equilibrium states.

One should understand that
\eqref{ef} represents a rather optimistic scenario. Formula \eqref{ef} wants to mimic
\eqref{efd} by replacing just one parameter. Why should
there be also out-of-equilibrium a single parameter and a useful notion of temperature in
its usual thermodynamic
 understanding, and how would it depend on
  the observables $V$ and $Q$? (See \cite{droz} for a very recent discussion.)
    Answers to these questions have
  been partially given but are often restricted within a context of mean
   field systems or for small fluctuations, effectively dealing with
calculations as in Appendix \ref{gau}, similar to calculations for scalar fields as in \cite{ckp}
and in \cite{gamb}. In fact, the optimism in \eqref{ef}
 is a sort of conservatism as it wants  to attach special reference
  to equilibrium forms and chooses to continue working with
   equilibrium notions such as the temperature.  We take a different attitude:
    the violation of
the equilibrium fluctuation-dissipation relation (FDR) is an opportunity to discover new
connections
between response and dynamical fluctuations away from equilibrium,
and to identify these relations in terms of the relevant newly emerging
physical quantities.

Let us then try to see how the notion of effective
temperature could be seen as a one-parameter reduction of a general
equilibrium-like FDR that is valid also outside equilibrium but with an effective dynamics.
The starting point is observing
that in the case of equilibrium, formula \eqref{efd} is equivalent with
\begin{equation}\label{efo}
  R_{QV}^{\rm{eq}}(t,s) \!=\!
   -\beta\langle (LV)(s)\,Q(t)\rangle_{\rm{eq}}
   \end{equation}
That follows from the calculation \eqref{eqfren}.  In other words,
the fluctuation-dissipation theorem in equilibrium can also be
called a fluctuation-frenesy theorem; the two terms on the
right-hand side of \eqref{form1} are simply the same.  Therefore,
for purposes of getting closer to equilibrium response formul{\ae}
one really has the choice to mimic either \eqref{efd} or rather
\eqref{efo}.  The first leads to the ambition of effective
temperature \eqref{ef}, the latter to the new notion  of effective
frenesy.  But the latter is also much richer.  In fact a simple
calculation, that we postpone to the end of Appendix \ref{per},
shows that the exact nonequilibrium response formula \eqref{form1}
can indeed be written in the equilibrium form \eqref{efo}:
\begin{equation}\label{inst}
 R_{QV}^\mu(t,s) =
  -\mean{G_{\mu_s}V(s)Q(t)}_{\mu}
\end{equation}
with a new effective frenesy
\begin{eqnarray}\label{Gs}
 G_{\mu}V &=& \frac{\beta}{2\mu}[L^{\dag}(\mu V)-VL^{\dag}\mu+\mu
   LV]\nonumber\\ &=&
 \frac{\beta\rho}{2\mu}[L^{*}(\frac{\mu}{\rho}
   V)-VL^{*}(\frac{\mu}{\rho})+\frac{\mu}{\rho} LV]
\end{eqnarray}
Here $L^{\dag}$ is the forward generator\footnote{It
is called the forward generator because of
the definition of the time-evolved measure $\mu_t$:
\[
 \mean{f(x_t)}_{\mu} = \int dx\, \mu_t(x)f(x)
 \]
From this we see that $\mu_t = e^{tL^{\dag}}\mu$, so that $L^{\dag}$
pushes the measure $\mu$ forward in time.
}
and for jump processes $L^{\dag} g(x) = \sum_y [W(y,x) g(y) -
W(x,y)g(x)]$, and $L^*$ is the adjoint generator, see
\eqref{adjoint} in Appendix \ref{per}. The first line in \eqref{Gs} does not
need to assume a stationary distribution $\rho$, while it is
explicitly and implicitly (in $L^*$) present in the second line of
\eqref{Gs}. At any rate the operator $G_{\mu}$ acting on $V$ in
\eqref{Gs} has the following exact property: it is itself a
generator (just like the original $L$ in \eqref{backward} or
\eqref{1st}) but of a new dynamics for which $\mu$ is an
equilibrium distribution (i.e. $\langle f\,(G_{\mu}g)\rangle_\mu =
\langle g\,(G_{\mu}f) \rangle_\mu$ for all $f,g$).  Thus, in
\eqref{inst} the generator $G_{\mu_s}$ is the instantaneous
equilibrium generator
 with respect to the time-evolved distribution $\mu_s$.
This underlies the conclusions of \cite{gaw2}.

   We compare \eqref{inst} with \eqref{efo} and we
recognize the equilibrium form with $L$ for the equilibrium $\rho$
replaced by $G_{\mu_s}$ for the transient $\mu_s$.  In the
stationary nonequilibrium case, we have $\mu_s=\rho$ and
\eqref{Gs} is
\begin{equation}\label{sa} G_\rho = \frac{\beta}{2} (L + L^*)
\end{equation}
 replacing $L$ in \eqref{efo}. Hence, if the perturbation $V$ is
time-direction independent in the precise sense that $LV=L^*V$,
 then the nonequilibrium response  (\eqref{st} = \eqref{form1} for  $\mu=\rho$)
  reduces to the equilibrium formula \eqref{efd} and $X=1$.
See  \cite{sasa} for very related conjectures and observations.

\section{Some other previous formulations}\label{prev}

The literature on (extensions of) the fluctuation-dissipation
theorem is vast. We apologize for not being able to list all
possible and even essential contributions. The equilibrium
formulation spans all of the previous century, while
nonequilibrium versions started to appear since the 1970's and
very much continue up to now. Early works include \cite{ag,wei}
and also \cite{weder,hanggi}, where we see a discussion within the
theory of stochastic dynamics.  In contrast to \cite{hanggi,ha}
our unperturbed process is time-homogeneous. Coming to more recent
times, violations of equilibrium FDR have been most often
discussed in transient regimes.  For example, in the context of
ageing phenomena \cite{henk,cr}, much thought has been given to
making sense of an effective temperature as briefly discussed in
Section \ref{efft}.

  However, recently much new
work has been also directed to find generic extensions of FDR in
nonequilibrium steady states and to discussions of the dissipative
elements in relaxations to nonequilibrium. There is for example
the line starting from \cite{house,house2,sasa,har} which treats
nonequilibrium heat effects.  In particular \cite{har} might also
be useful in real experiments because the FDR violation is there
directly connected with the energy dissipation.  We do not know
yet how to relate that to the new ideas surrounding the frenetic
term in our work or to the physics of ``active states'' as also in
\cite{ru}.  The latter follows the line of \cite{gg}, where the
connection between the so called fluctuation theorem and the
fluctuation-dissipation theorem was first explained.

  For other recent extensions of the
FDR, we refer to \cite{vulp,jona,ss,ss1,blickle}. We also
mentioned before (at the end of Section \ref{efft} and more will
come in Appendix \ref{per}) how our approach is related to the
co-moving frame interpretation of \cite{gaw,cili,gaw2}. There, as
in the previous references, one disadvantage is that one keeps the
stationary density $\rho$ (or its logarithm) as observable in the
fluctuation formula. In our approach the largely unknown
distribution only enters in the statistical averaging. Being
optimistic one could say that the generalized FDR of
\cite{vulp,ma} can be used to verify hypothesis on the phase space
distribution function.  Or, one can imagine approximation schemes
for making the numerics possible at all, cf. \cite{ss1}. In the
same sense the extensions in \cite{jona,gaw} are also not explicit
as they need the information on the adjoint hydrodynamics or again
on the local probability current, but at least they come with a
new and still interesting interpretation (reverse response in
\cite{jona} and co-moving frame in
\cite{gaw,gaw2}).

A first generalized fluctuation-dissipation relation giving a
response formula {\it in our sense} appears in Section 2 of
\cite{ckp}. It treats a Langevin dynamics $y(t)$  for soft spin
models and the resulting equation (2.10) in \cite{ckp} is exactly
our formula \eqref{asymm} for $Q=V$ equal to
$y$. A very similar treatment is repeated in \cite{lipp} for
systems of Ising spins (i.e., within the class of Markov jump
processes). 
Some more general treatment again for jump
processes is offered in \cite{diez}, in particular its equations
(16)-(17). However as before and as mentioned already in the
abstract of \cite{diez}, `` the asymmetry... is not related to any
physical observable.'' In contrast, we have emphasized the
interpretation via dynamical activity/frenesy in nonequilibrium
fluctuation theory. In fact, that interpretation is exactly what
makes a systematic generalization possible at all, as will become
even more clear in \cite{part2}.  The only study in which we
recognize some of the ideas related to the frenetic term is in
\cite{ru}, in the context of dynamical systems.

\section{Conclusions}
We have studied linear response relations under general
nonequilibrium conditions (stationary and not).
This first part has dealt with jump
processes and with overdamped diffusions, which is the usual
set-up for discussions on the violation of the
fluctuation-dissipation theorem and for the possible emergence of
an effective temperature. Here we have stressed the emergence of
the dynamical activity, or frenesy in linear order, as
complementary to the entropy flux. Out of equilibrium, dissipation
and activity detach and the response needs
 to be evaluated in terms of both entropic and frenetic correlation functions.
  As both correlations are expressed in
terms of explicit averages, they constitute a formula ready to use
in a general context. For example, estimates of these correlations
can be obtained with usual averaging in simulations, without any
need to know further details or approximations of the stationary
density of states. On the theoretical side, several previous
approaches are recovered or have been extended within the same
scheme. In many cases they have been discussed in specific model
dynamics or for specific observables. It is interesting that there
is a unifying approach w\`{\i}th statistical interpretation behind
this very broad variety of previous results.\\

Within the models we considered, more general types of perturbation
are possible and interesting. For example, what will happen if the perturbation
is not of the potential type but actually changes the nonequilibrium part or driving of the
dynamics? In all these cases our results must be modified, of course, but
we think that the general method and framework we use is applicable to this wider set of
questions. Specifically, the concepts of entropy and frenesy will remain the major
actors in nonequilibrium fluctuation-response theory.

 \vspace{1cm}

\noindent {\bf Acknowledgments:} The authors are grateful to
Wojciech De Roeck, Andrea Gambassi, Krzysztof Gaw\c{e}dzki, Dragi
Karevski and Karel Neto\v{c}n\'{y} for very useful discussions.
B.W. receives support from FWO, Flanders. M.B. benefits from
K.U.Leuven grant OT/07/034A.

 \vspace{1cm}

\appendix

\section{Explicit calculations for linear diffusions}\label{gau}
We give here the results of some explicit calculations to illustrate
 formula \eqref{x} in Section \ref{efft}.  Similar calculations can be found
  in \cite{ckp,gamb}.

Fix parameters $\alpha, B \in \bbR, D>0$ and look at the linear
Langevin dynamics for a global order parameter $M \in \bbR$,
\[
\dot{M}(t) = -\alpha\,M(t) + h_t\,B + \sqrt{2D}\,\xi(t)
\]
for standard white noise $\xi(t)$. The $h_t, t>0,$ is a small time-dependent field. The
generator of the unperturbed dynamics (on observables $f$) is
$Lf(M) = -\alpha \,M\, f'(M) + D\,f''(M)$.  We can think
 of a Gaussian approximation to a relaxational dynamics
of the scalar magnetization $M$ (no conservation laws and no spatial
structure) valid in high enough dimensions (above $d=4$ for the standard Ising model). Then,
 in a way, $\alpha = 0$  corresponds to the
critical (massless) dynamics and $\alpha >0$ is a paramagnetic dynamics (high temperature).
By taking $D\downarrow 0$ we exclude the diffusive aspects and we can think then of gradient
   relaxation in the low temperature regime.

The equilibrium (reversible stationary density on $\bbR$ for perturbation $B=0$) is
\[
\rho(M) =\frac 1{Z}\exp \left\{-\alpha\frac{M^2}{2D}\right\}
\]
with zero mean and variance $\langle M^2\rangle = D/\alpha$. (Here, $\alpha > 0$ is needed but
not for the
finite-time existence of the dynamics.)

We now start from an initial fixed $M(t=0) = M_0$.
The response function is obtained from
\[
\langle M(t)\rangle^h_{M_0} = M_0\,e^{-\alpha t} + B\int_0^t\id s\,h_s\,e^{-\alpha(t-s)}
\]
or
\begin{equation}\label{res}
\frac{\delta}{\delta h_s}\langle M(t) \rangle^h_{M_0}(h=0) = B\,e^{-\alpha (t-s)}
\end{equation}
which does in fact not depend on $M_0$ (and thus also equals the equilibrium result).

The correlation function for $0<s<t$ is
\[
\langle M(s)\,M(t)\rangle_{M_0}= M_0^2\,e^{-\alpha(t+s)} + \frac{D}{\alpha}
\big[ e^{-\alpha(t-s)} - e^{-\alpha(t+s)}\big]
\]
and hence
\[
\frac{\id}{\id s}\langle M(s)\,M(t)\rangle_{M_0} = -\alpha M_0^2\,e^{-\alpha(t+s)} + D
\big[ e^{-\alpha(t-s)} + e^{-\alpha(t+s)}\big]
\]
When we average that last expression over the equilibrium density (thus replacing $M_0^2$ by
$D/\alpha$) we find
\[
\frac{\id}{\id s}\langle M(s)\,M(t)\rangle_{\rho} =  D\, e^{-\alpha(t-s)}
\]
which, in comparison with \eqref{res} specifies the equilibrium temperature to be equal to
$T=D/k_B$.

The frenetic term is  obtained from $LM= -\alpha M$, and thus
\[
\langle LM(s)\,M(t)\rangle_{M_0} =
-\alpha\,M_0^2\,e^{-\alpha(t+s)} - D
\big[ e^{-\alpha(t-s)} - e^{-\alpha(t+s)}\big]
\]
Clearly,
\[
\frac{\delta}{\delta h_s}\langle M(t) \rangle^h_{M_0}(h=0)
=\frac{B}{2D}\{\frac{\id}{\id
s}\langle M(s)\,M(t)\rangle_{M_0}-\langle LM(s)\,M(t)\rangle_{M_0}\}
\]
as it should.

For the issue of effective temperature we compute the ratio ``frenesy versus
entropy''
as
\[
Y = Y(M_0;s,t) =
\frac{-\alpha \langle M(s) M(t)\rangle_{M_0}}{\frac{\id}{\id s}\langle
M(s)M(t)\rangle_{M_0}} =
\frac{-\alpha M_0^2e^{-\alpha(t+s)} - D \big[ e^{-\alpha(t-s)} -
e^{-\alpha(t+s)}\big]}{-\alpha M_0^2e^{-\alpha(t+s)} + D \big[ e^{-\alpha(t-s)} +
e^{-\alpha(t+s)}\big]}
\]
In that notation, the effective inverse temperature is $T^{\mbox{eff}} = 2T/(1-Y)$. In
equilibrium
$Y=-1$ while $Y=1$ for $D=0$ and $M_0\neq 0$.

For the limit of the ``frenetic ratio''
\begin{eqnarray}
\lim_{s\uparrow+\infty}\lim_{t\uparrow +\infty} Y(M_0;s,t) &&= -1, \quad \mbox{ if
}
\alpha>0\nonumber\\
&& = 0,  \quad \mbox{ if } \alpha=0\nonumber
\end{eqnarray}
with respectively, $T^{\mbox{eff}} = T$ (paramagnetic) and $T^{\mbox{eff}} = 2T$ (critical
quench), cf. \cite{gamb}.

\section{Discrete time}\label{disc}
So far we have been dealing with continuous time Markov processes.
Nevertheless simulations often use an updating in discrete time.
It is therefore useful to give the response formula also for discrete time,
here in analogy with the continuous time jump processes of
Section~\ref{jum}.  Similar efforts have appeared before in
\cite{Xa,lipp}.

Consider a discrete time  Markov process with configurations
$x,y,...$ and transition probabilities $p(x,y)$. We perturb the
system as follows: the new transition probability at time $i$ reads
\[ p^{h}_i(x,y) = \frac{p(x,y)}{z_i(x)}e^{\frac{h_i\beta}{2}[V(y)-V(x)]} \]
with normalization
\[ z_i(x) =  \sum_yp(x,y)e^{\frac{h_i\beta}{2}[V(y)-V(x)]}\]
Now consider a path $\omega = (x_i)$, with $i=1,\ldots,n$, where
at time $i$ the system jumps to configuration $x_{i+1}$. Then the
relative  probability of the path in the perturbed versus the
unperturbed dynamics is
\[ \frac{P^h}{P}(\omega) = \exp\left\{ \frac{\beta}{2}
\sum_{i=1}^{n-1}h_i[V(x_{i+1})-V(x_i)] -
\sum_{i=1}^{n-1}\log z(x_i) \right\} \] The response
function becomes
\begin{eqnarray*}
R_{Q,V}(n,m) &=& \left.\frac{\partial}{\partial h_m}\left< Q(x_n)
\right>^h_{\mu}\right|_{h=0}\\
&=& \frac{\beta}{2}\left<V(x_{m+1})Q(x_n)\right>_{\mu} -
\frac{\beta}{2}\left<\sum_yp(x_m,y)V(y)Q(x_n)\right>_{\mu}, \quad
m=1,\ldots,n-1
\end{eqnarray*}
Again we recognize this as the sum of an entropic term
\[
\frac{\beta}{2}\left<[V(x_{m+1})-V(x_m)]Q(x_n)\right>_{\mu}
 \] and a frenetic term
\[
 \frac{\beta}{2}\left<\sum_yp(x_m,y)[V(y)-V(x_m)]Q(x_n)\right>_{\mu}
 \]
In equilibrium $\rho$, when the process is time-reversal symmetric, that simplifies to
($m<n$)
\[ R_{Q,V}^{\rm{eq}}(n,m) = \frac{\beta}{2}\left<[V(x_{m+1})-
V(x_{m-1})]Q(x_n)\right>_{\rm{eq}}
\]
Note that in this discrete version, the entropic and the frenetic
terms do not give exactly the same contribution to the equilibrium
formula,  although this discrepancy disappears in the limit to
continuous time.

\section{First order perturbation}\label{per}

The usual and natural approach to response theory is that of
time-dependent perturbation theory, see for example \cite{mar}.
We show here what that gives for the overdamped diffusion
processes of Section \ref{overd}.

We prepare the system at time $t=0$ according to its stationary
distribution $\rho$; the perturbation $-h\,V$ is added for positive times. Therefore, for times
$t\geq 0$ the
dynamics has (backward) generator (working on observables)
\[
L^h = L + h \,\nu\nabla V \cdot \nabla,\quad L = \nu(F-\nabla U
+ \nabla D)\cdot\nabla
\]
with $\nu = \beta\,D$.
For the change in
expectations at times $t$ with respect to what we had at time
zero
\[
\langle Q(t)\rangle^h - \langle Q(0)\rangle = \int \id x\,
\rho(x)\,\big(e^{tL^h} - e^{tL}\big)Q(x)
\]
we get the linear order
\[
e^{tL^h} - e^{tL} = \int_0^t e^{(t-s)L}\,(L^h-L)\,e^{sL} \id s +
O(h^2)
\]
Or, always to leading order in $h\downarrow 0$,
\[
\frac{1}{h}\big[\langle Q(t)\rangle^h - \langle Q(0)\rangle\big]
=\int_0^t\id s\,R_{QV}(t,s)
\]
with response function
\begin{equation}\label{dys}
R_{QV}(t,s) = \int \id x\, \rho(x)\,\nu\nabla V(x)\cdot \nabla
e^{(t-s)L}Q(x)
\end{equation}
(We are still writing the dependence separately on time $s$ and on
time $t$, for greater generality in case the perturbation is
time-dependent (through $h_s$).)  The equation \eqref{dys} is not useful as such
because the derivatives do not commute with the time-evolution. Another way would be trying the
partial integration
\[
\nabla\cdot\big(\rho\,\nu\nabla V\big) =  \nabla \rho\cdot
\nu\nabla V + \rho\,\nabla\cdot\nu\nabla V =\beta\rho\{\,LV -
\frac{j_\rho}{\rho}\cdot\nabla V\}
\]
where we have inserted the local stationary current $j_\rho$
(assuming that $\rho$ nowhere vanishes). As a result
\begin{equation}\label{dys1}
R_{QV}(t,s) = -\beta \int \id x\, \rho(x)\,{\cal L} V(x)\,
e^{(t-s)L}Q(x)\quad \mbox{ for } {\cal L} = L - u(x)\cdot\nabla
\end{equation}
with local velocity
\[ u(x) = \frac{j_\rho}{\rho}(x)
\]
function of the stationary density $\rho$.

  We can further rewrite that using the
  adjoint generator $L^*$ with respect to $\rho$: for any pair of functions $f,g$
\begin{equation}\label{adjoint}
 \int dx \rho(x)g(x)L^*f(x) = \int dx \rho(x)f(x)Lg(x)
\end{equation}
Its physical interpretation is that it generates the time-reversed stationary process.
It is important to realize though that this adjoint generator $L^*$ depends on
the stationary distribution $\rho$, and is therefore most often not explicitly known in
nonequilibrium systems. For equilibrium where $\rho\propto \exp (-\beta U), L=L^*$.

We can verify that $L- L^* = 2 u\cdot\nabla$ so that ${\cal L} =
L^* + u\cdot\nabla$ in \eqref{dys1}:
\begin{equation}\label{dys2}
R_{QV}(t,s) = \beta \frac{\id}{\id s}\int \id x\, \rho(x)\,
V(x)\, e^{(t-s)L}Q(x) -  \beta\int \id x\, \rho(x)\, u(x) \cdot
\nabla V(x)\, e^{(t-s)L}Q(x)
\end{equation}
For jump processes, under Section \ref{jum},
 \begin{eqnarray*}
((L- L^*)V)(x) &=& 2\sum_y\frac{j(x,y)}{\rho(x)}\,[V(y)-V(x)]\\
j(x,y) &=& W(x,y) \rho(x) - W(y,x) \rho(y)
\end{eqnarray*}
The equations \eqref{dys1}--\eqref{dys2} contain the
interpretation that the equilibrium form gets ``restored'' when
describing the system in the Lagrangian frame moving with drift
velocity $u$, as in \cite{gaw}.  Note that ${\cal L} = L - L_A$, which subtracts
the antisymmetric part $L_A = (L-L^*)/2$ from the original generator $L=L_S + L_A$. What remains
is
the symmetric part ${\cal L}=L_S$, of course defining
an evolution which is now detailed balance with respect to $\rho$. That aspect can also be realized
path-wise:
when $\Phi_t(x)$ is the one-parameter group corresponding to the flow with velocity $u(x)$,
\[
\partial_t\Phi_t(x) = u(\Phi_t(x))
\]
then, $y_t = \Phi_{-t}(x_t)$ satisfies an equilibrium Langevin
equation with potential $-\beta^{-1}\ln \rho$  but with
time-dependent coefficients (in mobility).  In other words, the
passage to the Lagrangian frame of local velocity removes the
non-conservative forcing, as explained in \cite{gaw,gaw2}.
  Still, if we do not know
$\rho$, the formul{\ae} \eqref{dys1}--\eqref{dys2} contain unknown
observables for statistical averaging. Moreover, thinking of spatial processes, the
probability current has little relation with the real physical
currents. For our formula \eqref{form1}, one has an explicit
expression in terms of known observables and $\rho$ only enters
the statistical averaging.

Let us now see how the above expressions are mathematically related to our formula \eqref{form1}.
For stationary nonequilibrium,
taking $\mu=\rho$, it is immediate to rewrite \eqref{form1} as
 \begin{equation}\label{st}
   R_{QV}(t,s) \!=\!
   \,\beta\frac{\id}{\id s}\!\left< V(s)Q(t)\right>_\rho
 - \frac{\beta}{2}\left<\big(L\!-\!\!L^*\big)V(s)\,Q(t)\right>_\rho
 \end{equation}
  In fact,
 \begin{eqnarray}\label{rev}
\frac{\id}{\id s}\!\left< V(s)Q(t)\right>_\rho &&
= -\frac{\id}{\id t}\!\left< V(s)Q(t)\right>_\rho \nonumber\\
&&= -\langle (L^*V)(s)\,Q(t)\rangle_\rho
\end{eqnarray}
so that also
\begin{equation}\label{sym}
  R_{QV}(t,s) \!=\!
   -\beta\left< \left(\frac{L+L^*}{2}V\right)(s)\,Q(t)\right>_\rho
   \end{equation}
as we have also seen in \eqref{sa}, or in \eqref{dys1} because ${\cal L}
= (L + L^*)/2$.

Also for transient regimes we  can rewrite the response formula
\eqref{form1} as what one would get in equilibrium plus a
correction term:
\begin{equation}\label{tr}
 R_{QV}^\mu(t,s) = \beta\frac{\id }{\id s}\mean{V(s)Q(t)}_{\mu} +
 \beta\mean{\tilde{G}_{\mu_s}V(s)Q(t)}_{\mu}
\end{equation}
where now
\[ \tilde{G}_{\mu}V = \frac{1}{2\mu}[-VL^{\dag}\mu+L^{\dag}(\mu V)-\mu LV] \]
Formula \eqref{tr} reduces to \eqref{st} for $\mu=\rho$.

\bibliographystyle{plain}

\end{document}